# Self-mixing interference in a thin-slice solid-state laser with few feedback photons per observation period

Kenju Otsuka[1*] and Seiichi Sudo[2]

[1]*TS³L Research, Yamaguchi 126-7, Tokorozawa, Saitama 359-1145 Japan*
[2]*Department of Physics, Tokyo City University, Tokyo 158-8557 Japan*
*Corresponding author: kenju.otsuka@gmail.com

Formation of a peculiar lasing pattern leading to a $TEM_{00}$-like mode accompanied by a nonzero bright outer-ring emission (namely, a modified $TEM_{00}$ mode) was observed in a 300-μm-thick $LiNdP_4O_{12}$ (LNP) laser with reflective flat end mirrors that was pumped with a laser diode whose focused spot size was larger than the lasing beam spot size determined by the thermal lens effect. The photon lifetimes of the laser oscillations were found to be greatly shortened to $\tau_p$ = 6.74 ps as compared with those in pure $TEM_{00}$ mode operation, while the fluorescence lifetime was $\tau$ = 130 μs. The formation of the peculiar transverse mode was theoretically treated in terms of the pump-intensity-dependent thermal lens effect, which yields the lasing beam spot sizes in the resultant optical cavity. In addition, transverse spatial hole burning of population inversions due to the preceding transverse eigenmode intensity and the associated anti-guidance of the pure $TEM_{00}$ field leading to formation of the $TEM_{00}$ with an emergent outer ring emission was shown to appear with increasing pump power through the electric lens (i.e., population lens) effect, in which almost all of the population inversion contributes to lasing with a pronounced effective modal gain. Self-mixing laser Doppler velocimetry (LDV) experiments and numerical simulations revealed that spontaneous emission factors in the peculiar mode oscillations were greatly decreased in comparison with those in pure $TEM_{00}$ mode operation. The present thin-slice solid-state laser possessing a large fluorescence-to-photon lifetime and decreased spontaneous emission factor enabled us to detect LDV signals in the regime of few feedback photons per observation period.

**PhySH:** Laser dynamics, Optical pumping, Spatial profiles of optical beam, Optical interferometry, Spontaneous emission, Photon statistics, Optical chaos

## I. INTRODUCTION

Kogelnik and Li published a monumental work on optical resonators based on the principles of optics and found that Hermite-Gaussian modes form as stable orthogonal transverse eigenmodes in Fabry-Perot optical cavities [1]. Additionally, Fabry-Perot microcavities, such as vertical-cavity surface-emitting laser diodes (VCSELs) and laser-diode (LD)-pumped thin-slice solid-state lasers with coated end mirrors (abbreviated as TS³Ls), have been revisited in a different context related to the spatiotemporal dynamics of transverse modes in Fabry-Perot microcavities [2].

The Fresnel number of a thin-platelet TS³L cavity, $N_F = a^2/l\lambda_o$ ($a$: aperture radius, $l$: optical cavity length, $\lambda_o$: lasing wavelength), is on the order of $10^2 - 10^3$ larger than those of conventional cavities. This allows light to propagate in the laser cavity with inherently low diffraction loss in accordance with the plane-wave approximation. Moreover, it enables lasing in a variety of transverse modes to take place depending on the shape and spot size of the pump beam, i.e., by controlling the gain as well as the thermally-induced refractive index confinement of the lasing transverse modes. These forms of lasing include vortex arrays originating from a higher-order Ince–Gauss mode in elliptical coordinates and rectangular-type vortex arrays born from Hermite-Gauss modes in a 300-μm-thick $LiNdP_4O_{12}$ (LNP) laser with shaped wide-aperture LD end pumping [3]. In particular, it has been found that the pure $TEM_{00}$ mode oscillation dominates higher-order transverse modes for a TS³L platelet laser with coated end mirrors when a circular pump beam is tightly focused with an objective lens through the enhanced thermal lens effect [4].

In the study reported here, we demonstrated that a peculiar modified $TEM_{00}$ mode operation with outer-ring emissions takes place in a 300-μm-thick $LiNdP_4O_{12}$ (LNP) laser by controlling the thermal lensing, i.e., by conducting heat away from the pump-beam focus, under wide-aperture LD end pumping. We found that the slope efficiency is high due to the enhanced modal gain and that a unique merge-emerge process of outer-ring emissions arises from the combined effect of pump-dependent thermal lensing and population lensing associated with transverse spatial hole burning. Moreover, the modified $TEM_{00}$ mode laser was shown to possess an extremely short photon lifetime.

In addition, we carried out ultra-high-sensitivity laser Doppler velocimetry (LDV) experiments with the modified $TEM_{00}$ mode in the regime of few feedback photons per observation period. As background, laser feedback interferometry [5] as well as highly sensitive self-mixing intensity modulations have been reported in class-B lasers, e.g., in a thin-slice LNP laser [6] and in a laser diode [7], subjected to weak Doppler-shifted optical feedback from a rotating object. While traditional laser Doppler velocimetry (LDV) systems utilize the Mach-Zehnder heterodyne interference between the lasing light and Doppler-shifted light from a moving target, self-mixing LDV works through the intensity modulation effect of a laser due to the interference between the lasing and feedback fields. In short, the self-mixing laser acts both as a mixer-oscillator and highly sensitive detector for a signal from the target. The resultant optical sensitivity has been shown to be enhanced in proportion to the square of the fluorescence-to-photon lifetime ratio, $K = \tau/\tau_p$, where the K value of state-of-the art $TS^3Ls$ reaches the order of $10^5 – 10^6$ [8, 9]. In fact, many applications using $TS^3Ls$ have been demonstrated, including displacement, ranging, vibrometry at nanometer resolution [10-12], and imaging of an object in turbid media [8, 13-15]. The short cavity lengths of $TS^3Ls$ ensure a short photon lifetime, but they also increase the laser's intrinsic quantum (spontaneous emission) noise because the decreased active volume increases the spontaneous emission coupling factor, β [16]. This implies that the signal-to-noise ratio (SNR) is not always improved by reducing the cavity thickness.

We realized that a significant improvement in optical sensitivity could be expected in self-mixing LDV by utilizing the modified $TEM_{00}$ mode $TS^3L$ because of its extremely short photon lifetime, $\tau_p = 6.74$ ps (i.e., $K = 1.93 \times 10^7$). Our ultra-high sensitivity LDV experiments with the modified $TEM_{00}$ mode in the regime of few feedback photons per observation period revealed a significant enhancement in SNR in comparison with state-of-art self-mixing LDV systems, reflecting the decreased spontaneous emission coupling factor, $\beta = 10^{-9}$, associated with the wide-aperture pumping, together with the extremely large K value.

The paper is organized as follows: Section II describes the structural changes in the transverse modes that depend on the spot size of the pump beam and the basic properties of these modes, e.g., slope efficiency and photon lifetime. It also provides theoretical results on thermal and population lens effects associated with transverse spatial hole burning of population inversions, which reproduce the experimental results. Section III is devoted to the pump-dependent merge-emerge process of outer-ring emissions in the modified $TEM_{00}$ mode and features a supplementary movie on the structural stability.

Section IV describes the high-sensitivity and low-noise self-mixing LDV experiments that we conducted on the modified $TEM_{00}$ laser. It discusses the significant improvement in SNR as compared with the $TEM_{00}$ mode operation as well as the self-mixing interference in the regime of a few feedback photons. An intriguing analogy with photon statistics is shown to exist wherein the Poisson distribution approaches a Gaussian as the average number of feedback photons per observation period increases.

## II. STRUCTURAL CHANGE IN TRANSVERSE MODE WITH PUMP-BEAM SPOT SIZE

### A. Experimental setup

The experimental setup is shown in Fig. 1(a). A nearly collimated lasing beam from a laser diode (wavelength: 808 nm) was passed through an anamorphic prism pair to transform an elliptical beam into a circular one, and it was focused onto a thin-slice laser crystal by a microscopic objective lens with a numerical aperture, NA, of 0.5 and an absorption coefficient at 808 nm is $\alpha_P = 115$ cm$^{-1}$. The pump beam profiles are depicted in the inset. The laser crystal was 5-mm-diameter clear-aperture, 300-μm-thick direct compound $LiNdP_4O_{12}$ (LNP) (pseudo-orthorhombic b-plate); a dichroic mirror $M_1$ (transmission at 808 nm > 95%; reflectance at $\lambda = 1048$ nm, $R_1 = 99.8\%$) was coated on one of the end surfaces and a flat mirror $M_2$ (reflectance at $\lambda = 1048$ nm, $R_2 = 99\%$) was attached to the crystal. The detailed cross-sectional view of the cavity is shown in the inset of Fig. 1(a). The high Nd concentration, about 30 times higher than in doped laser crystals, results in an extremely short absorption length of $1/\alpha_P = 87$ μm for 808-nm pump light [17]. This implies that the pumped region is localized at 87 μm in depth from the crystal surface coated by mirror $M_1$ and that heat easily dissipates to heat sinks directly in contact with the crystal or through the output mirror in contact with the heat-sink compound. In the conventional $TS^3L$ with coated end mirrors depicted in Fig. 1(b), the crystal surface $M_2$ is in contact with the copper plate and strong transverse mode confinement is brought about through the enhanced thermal lens effect due to the weak heat dissipation from the localized pumped region. In this case, $TEM_{00}$ mode oscillations occur independently of the pump spot size [4].

Near- and far-field lasing patterns were measured by a PbS phototube followed by a TV monitor. Lasing optical spectra were measured with a scanning Fabry-Perot interferometer (Burleigh SA$^{PLUS}$; 2 GHz free spectral range; 6.6 MHz resolution). Linearly polarized single frequency emissions along the c-axis were observed, reflecting the fluorescence anisotropy.

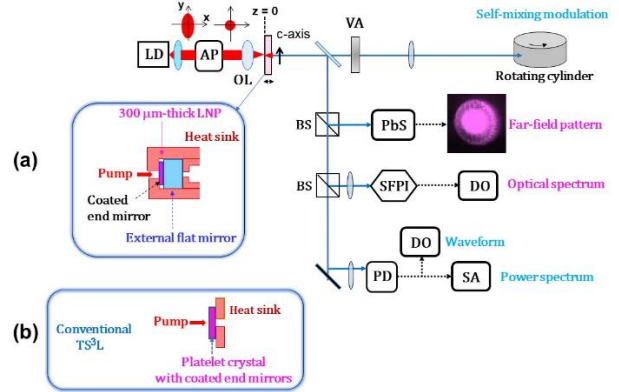

FIG. 1. (a) Experimental apparatus. PD: photo-diode, AP: anamorphic prism pair. SFPI: scanning Fabry-Perot interferometer, DO: digital oscilloscope, SA: spectrum analyzer, VA: variable optical attenuator, BS: beam splitter. (b) Cross-sectional view of conventional $TS^3Ls$ with coated end mirrors.

Figure 1(a) also illustrates the self-mixing laser Doppler velocimetry (LDV) scheme that was used to evaluate the optical sensitivity of the peculiar modified $TEM_{00}$ mode against tiny scattered fields from a moving target and the signal-to-noise ratio in comparison with the pure $TEM_{00}$ mode. Dynamic properties were examined by using an InGaAs photodiode (New Focus 1811, DC-125 MHz) followed by a digital oscilloscope (Tektronix TDS 3052, DC-500 MHz) and a spectrum analyzer (Tektronix 2712) depending on the purpose.

### B. Pump-beam focus-dependent lasing transverse pattern

In the case of $TS^3Ls$ without an external optical cavity, a stable resonator condition is achieved through the thermally induced lensing effect [4], and the input-output characteristics as well as the transverse and longitudinal mode oscillation properties depend directly on the focusing conditions (e.g., spot size and shape) of the pump beam on the crystal due to the mode-matching between the pump and lasing mode profiles [18]. In the present experiment, the pump-beam diameter was changed by shifting the laser crystal along the z-axis, as depicted in Fig. 1. The pump spot size, $w_p$, increased as the laser crystal was shifted away from the pump-beam focus along the z-axis (i.e., z > 0). The pump spot size could be varied from $w_p = 10$ μm (z = 0) to 70 μm (z = 2 mm). Typical far-field patterns observed for different pump beam spot sizes are shown in Figs. 2(a)-2(c). Single-longitudinal-mode oscillations were observed at a wavelength $\lambda_o$ of 1048 nm. A detailed optical spectrum measured by the scanning Fabry-Perot interferometer, which indicates a single-longitudinal mode oscillation, is shown in Fig. 2(d). $TEM_{00}$ mode oscillations (b) and (c) accompanied by outer-ring emissions were achieved quite reproducibly by adjusting the pump beam spot size. The scenario of the pump-dependent structural changes is explained later in section **III** with the help of Supplementary video clips.

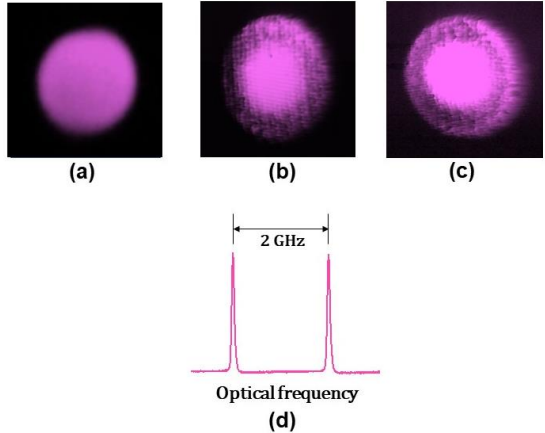

FIG. 2. Far-field patterns and optical spectrum. (a) Pure $TEM_{00}$ mode at pump spot size of $w_p = 10$ μm. (b), (c) Modified $TEM_{00}$ mode at $w_p = 52$ μm and 70 μm. (d) Optical spectrum indicating single-longitudinal mode operation.

The pure $TEM_{00}$ mode oscillation was obtained with a minimum threshold pump power of $P_{th} = 20$ mW at $w_p = 10$ μm (z = 0), where lasing spot size $w_o = 30$ μm. As $w_p$ was increased, the threshold pump power gradually increased and the resultant lasing transverse mode exhibited a structural change, as shown in Figs. 2(b) [$w_p = 52$ μm, $w_o = 43$ μm] and 2(c) [$w_p = 70$ μm, $w_o = 48$ μm]. For such a wide pump-beam focus, a ring-shaped optical field appeared surrounding the modified $TEM_{00}$ pattern, and the ring pattern became more intense as $w_p$ was increased up to 70 μm.

The input-output characteristics corresponding to the lasing patterns in Fig. 2 are shown in Fig. 3(a), where the blue, green, and red plots correspond to Fig. 2(a), (b) and (c), respectively. Surprisingly, the slope efficiencies, $S_e$, for modified $TEM_{00}$ mode operation were almost the same as that for the pure $TEM_{00}$ mode, $S_e = 9.5\%$, regardless of the increase in threshold absorbed pump power.

The pump-dependent relaxation oscillation frequency, which obeys the formula $f_{RO} = (1/2\pi)[(P/P_{th} - 1)/\tau\tau_p]^{1/2}$, was measured using the spectrum analyzer. The results are shown in Fig. 3(b). The photon lifetimes of the laser oscillator operating in these different transverse modes, $\tau_p$, were evaluated by using the above formula and the measured relaxation frequencies. The average photon lifetime became shorter with increasing pump beam spot size, from 70.2 ps (pure $TEM_{00}$) to 12 ps and 6.74 ps ($TEM_{00}$ featuring an outer ring).

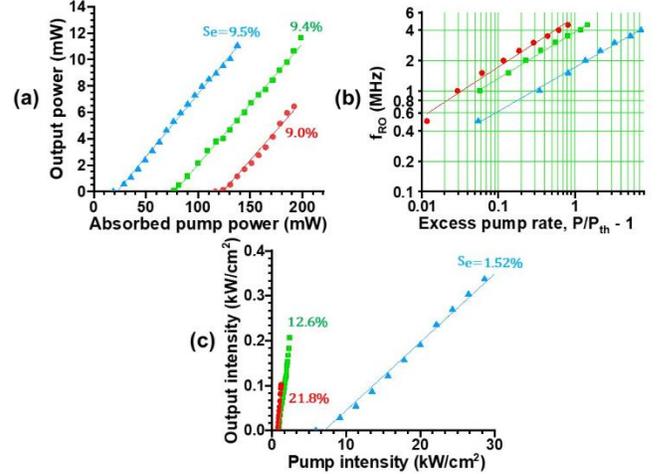

FIG. 3. (a) Input-output characteristics of pure and modified $TEM_{00}$ mode oscillations. (b) Pump-dependent relaxation oscillation frequencies. (c) Relation between input and output intensities. Triangle, square, and circle plots correspond to Fig. 2(a), (b), and (c), respectively.

The photon lifetime is related to the cavity loss, as follows:

$$\tau_p = -\frac{2n_0 l}{c[\ln\{R_1 R_2 (1-L)^2\}]}. \qquad (1)$$

Here, $n_0 = 1.6$ is the refractive index of LNP, $c$ is the light velocity, $l = 0.3$ mm is the cavity length, and $L$ is the cavity loss per passage excepting mirror transmission losses. The cavity losses were estimated from the measured photon lifetimes, as $L$ = 1.67%, 12 % and 20.7 % (Figs. 3(b)-3(d)).

Let us consider the physics behind the coincidence of the slope efficiencies for the pure and modified TEM$_{00}$ mode operations in spite of their totally different cavity losses. In the present study, the laser was operated under different pump-beam focusing conditions with narrow- and wide-aperture pump-beam focus. Therefore, the input-output characteristics in terms of the pump and lasing beam intensities, $I_p$ and $I_o$, are considered to be strongly modified. Results corresponding to Fig. 3(a) are shown in Fig. 3(c). As expected, the slope efficiencies for the TEM$_{00}$ mode with the outer-ring emission are greatly enhanced compared with that for the pure TEM$_{00}$ mode operation, mainly reflecting the difference between pump-beam spot sizes.

The circulating light intensity within the cavity and the resultant output power are given by

$$I_{cir} = I_s \left[ \frac{2G(I_p)l}{2L - \ln(R_1 R_2)} - 1 \right] = I_s \left[ \frac{I_p}{I_{p,th}} - 1 \right] \quad (2)$$

$$P_{out} = A(-\ln R_2) P_{cir} \quad (3)$$

Here, $I_s$ is the saturation intensity at the lasing wavelength, $I_p$ the pump intensity, $I_{p,th}$ the threshold pump intensity, $G(I_p)$ the modal gain coefficient, and $A = \pi w_o^2$ the effective cross section of the lasing beam interacting with the population inversion. G ($I_p$) depends on the pump-beam focusing conditions and the resultant interaction between the pump beam and lasing beam profiles, which in turn depends on the thermal lens effect in a complicated fashion. The circulating lasing intensity and the resultant output power are considered to be greatly increased, reflecting the pronounced slope efficiencies for the intensities shown in Fig. 3(c). This suggests that G ($I_p$) for the modified TEM$_{00}$ mode featuring outer-ring emissions is pronounced in spite of the larger cavity losses as compared with the pure TEM$_{00}$ mode, leading to the similar slope efficiencies shown in Fig. 3(a).

### C. Thermal lens effect in a thin-slice LNP laser with laser-diode end pumping

The thermal lens effect is the key to investigating the formation of a stable laser cavity that depends on the focusing conditions of the pump beam. Here, a theoretical analysis is performed for the LD-pumped LNP laser used in the experiments.

The thermal lens effect due to the radial refractive index change is brought about by the pump-dependent rise in temperature of the laser material associated with the nonradiative decay from the absorption state to the metastable state. The actual resonator in our experiment and the equivalent resonant cavity are shown in Fig. 4(a). The radius of curvature of a thermally-induced curved mirror, $R_T$, is given by [19]

$$R_T = \frac{\pi w_p^2 K_T}{Q \left( \frac{dn}{dT} \right)}, \quad (4)$$

$$Q = P \left( 1 - \frac{\nu_o}{\nu_p} \right). \quad (5)$$

Here, $w_p$ is the average pump spot size in the crystal, $K_T$ is the thermal conductivity, $dn/dT$ is the thermal-optic coefficient of the refractive index, $P$ is the absorbed pump power, and $\nu_o$ and $\nu_p$ are the lasing and pump optical frequencies. Q denotes the total heat generated by the quantum defect of the laser transition. Figure 4(b) shows $R_T$ calculated as a function of the pump spot size, $w_p$, depicted in Fig. 4(a), for absorbed pump powers, $P$, incremented in 20 mW steps, assuming the thermal constants of LNP, $K_T$ = 3.3 W/mK and $dn/dT$ = 7.1×10$^{-6}$/K [20].

The resultant spot sizes at the mirrors which form the hemi-spherical cavity are given by [1]

$$w_1 = \sqrt{\lambda_o/\pi} \sqrt[4]{n_0 l (R_T - n_0 l)}, \quad (6)$$

$$w_2 = \sqrt{\lambda_o R_T/\pi} \sqrt[4]{n_0 l/(R_T - n_0 l)} \quad (7)$$

The calculated spot sizes, $w_1$ and $w_2$, are plotted as a function of $R_T$ in Fig. 4 (c), assuming $n_0$ = 1.6 and $l$ = 0.3 mm. In addition, the pump-dependent spot sizes, $w_1, w_2$, are shown in Fig. 4(d) for pure and modified TEM$_{00}$ mode operations. The measured lasing spot sizes in experiment **II-B** are close to the theoretical values determined by Eqs. (4)-(7).

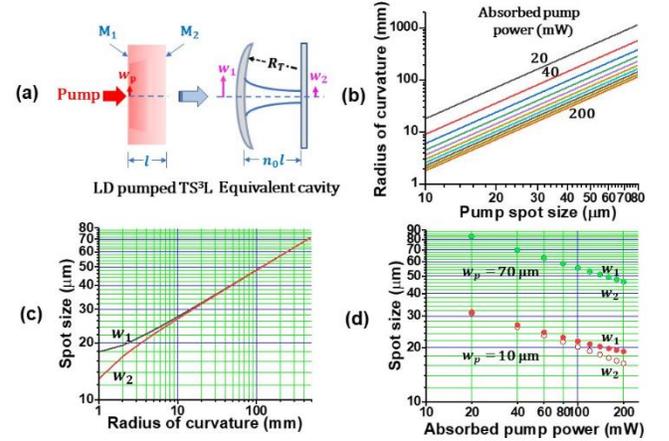

FIG. 4. (a) LD-pumped platelet LNP laser and its equivalent cavity. (b) Pump-dependent radius of curvature, $R_T$. (c) $R_T$-dependent spot sizes at cavity mirrors, $w_{1,2}$. (d) Pump-dependent spot sizes, $w_{1,2}$, for pure and modified TEM$_{00}$ mode operation with $w_p$ = 10 μm and 70 μm, respectively.

### D. Transverse spatial hole burning and population lens effect

In the case of $w_p > w_o$, the lasing modal field, whose beam cross section is sufficiently smaller than that of the pump beam, expends the population inversion density in the central part of the corresponding transverse eigenmode. Consequently, the remaining population inversions will surround the lasing mode if their spatial diffusion can be neglected. This effect is called transverse spatial hole burning and was investigated in anisotropic lasers operating in dual-polarization multi-transverse modes [21]. In this case, the higher-order transverse mode, which is orthogonally polarized to the former $TEM_{00}$ mode, tends to oscillate simultaneously by expending the remaining population inversions [21]. Spatial hole burning has also been investigated in oxide-confined VCSELs [22,23]

As for the $TEM_{00}$ mode with the surrounding ring-type far-field pattern that was observed when $w_p > w_o$, the transverse eigenmode field is expected to cause transverse spatial hole burning. The pump rate of the excited atoms and the decreasing rate at which atoms are excited by lasing photons through stimulated emission are given by $\sigma_P I_P/h\nu_P$ and $\sigma_e I_{cir}/h\nu_o$, respectively, where $I_P$ is the pump light intensity, and $\sigma_e$ and $\sigma_P$ are the emission and absorption cross sections for the pump beam, respectively [24].

Figure 5 shows the radial profiles of the remaining population inversion (i.e., the net excitation rate) corresponding to the $TEM_{00}$ mode operation featuring the outer-ring emission shown in Fig. 2(c) for various pump-dependent circulating intensities with different spot sizes, together with that for $I_{cir} = 0$, i.e., the threshold population inversion profile. Here, calculations were performed using Eq. (2), assuming $\sigma_e = 1.7\times10^{-19}$ cm$^2$ and $\sigma_P = 2.3\times10^{-19}$ cm$^2$ [25]. As the pump power (i.e., circulating light intensity) increases, depletion of the population inversions (i.e., transverse spatial hole burning) takes place.

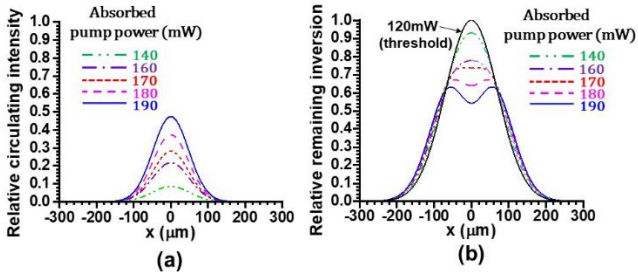

FIG. 5. Transverse spatial hole burning. (a) Profile of circulating intensity (b) Remaining population inversions in the presence of circulating intensities with different spot sizes. $w_p = 70$ μm. P = 140 mW: $w_o$ = 51 μm, 160 mW: 49.4 μm, 170 mW: 48.0 μm, 180 mW: 47.9 μm, 190 mW: 46.7 μm.

Here, the spatial integral of the remaining population inversion, $V = \iiint N(x,y,z)dv$, was found to coincide with the threshold value for $I_{cir} = 0$ for all absorbed pump powers within 4% error. This strongly implies that the population inversion density is kept at the threshold value under the lasing condition obeying laser theory. It should be noted that this basic rule is only valid for appropriate circulating laser intensities, which are estimated from the pump-dependent lasing spot sizes derived from Eqs. (4)-(7). This suggests that the effective modal gain coefficient $G(I_p)$, which determines the circulating light intensity given by Eq. (2), is enhanced for the modified $TEM_{00}$ mode to compensate its larger cavity loss, leading to it having similar slope efficiencies with the pure $TEM_{00}$ mode, as shown in Fig. 3(a). Indeed, the remaining population inversion shown in Fig. 5 is expected to give an additional effective gain for a noticeable radial emission, i.e., outer-ring emissions, around the main central modal field so that almost all of the population inversion contributes to lasing by wide-aperture pumping

On the other hand, the refractive index variation due to the difference in polarizability of the active ions in the metastable and ground states, $\Delta\alpha = \alpha_e - \alpha_g$, is considered to increase the cavity loss, $L_T$, through depletion of the population inversions, i.e., the defocusing effect. This effect is expected to be pronounced in direct compound Nd laser materials like LNP having a high total ion density of $N_T = 4.37 \times 10^{21}$ cm$^{-3}$ [17]. Indeed, the population-inversion-dependent refractive index variation, $\Delta n$, can be expressed as [26, 27]

$$\Delta n = \left(\frac{2\pi}{n_0}\right)f_L^2 N_e \Delta\alpha, \qquad (8)$$

where $f_L = (n_0^2 + 2)/3$ is the Lorentz local-field correction factor, and $N_e$ denotes the excited-state ion population. Usually, $N_e$ is, to first order, proportional to the pump intensity, $N_e \approx N_T(I_P/I_{s,a})$ (for $I_P \ll I_{s,a}$), where $I_{s,a} = hc/\lambda_e \sigma_P \tau$ is the saturation intensity at the excitation wavelength. Since $N_e$ is proportional to $I_P$, Equation (8) can be written in terms of the pump-intensity dependent refractive index change, $\Delta n = n_2' I_P$: [26, 27]

$$n_2' = (2\pi/n_0)f_L^2 N_T \Delta\alpha/I_{s,a}. \qquad (9)$$

For LNP crystals, the value of $\Delta\alpha$ is unknown, but their high Nd-ion concentration is considered to result in the defocusing effect on the central transverse mode field arising from the wide-aperture pumping and an increase in the cavity loss, e.g., diffraction loss. Moreover, a pronounced modal gain $G(I_p)$ and refractive index confinement following Eq. (9) for the outer-ring emission are expected to be associated with modified $TEM_{00}$ mode oscillations.

### III. PUMP-DEPENDENT STRUCTURAL CHANGE OF MODIFIED $TEM_{00}$ MODE

The previous section examined the modified $TEM_{00}$ mode oscillations under wide-aperture laser-diode end pumping and discussed the physical origin of the non-zero ring emission surrounding the $TEM_{00}$ main lasing mode in terms of competing processes of thermal-lens-mediated optical confinement by the pump beam and the anti-guidance effect due to population lensing. The outer-ring emissions were interpreted as being due to anti-guidance of the main modal field by population lensing, which follows the remaining population inversion created through the transverse spatial hole burning by the main mode, while they are expected to get some amplification gain from the remaining population inversion surrounding the main mode.

An intriguing question arises as to whether such outer-ring emissions are structurally stable in the sense that they maintain their shape with increasing pump power as normal Hermite–Gauss and Laguerre–Gauss modes do. Motivated by this question, we examined pump-dependent structural changes in the near-field as well as far-field. Typical examples corresponding to Fig. 2(c) are shown in Fig. 6.

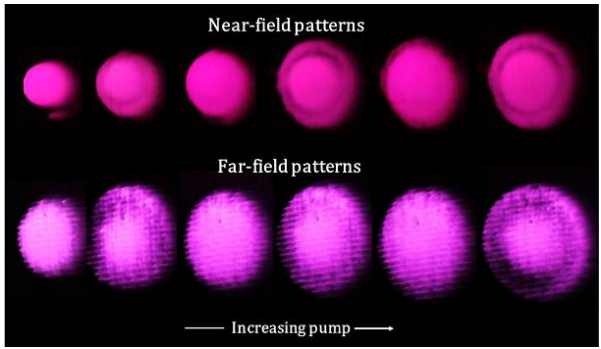

FIG. 6. Pump-dependent structural change in the modified $TEM_{00}$ mode corresponding to Fig. 2(c). The pump spot size, $w_p$, was 70 μm. The absorbed pump power was varied from 140 mW to 300 mW. See the Supplementary movies.

The observed structural changes in the modified $TEM_{00}$ transverse mode follow essentially the same scenario for both the near- and far-field patterns and are rooted in the physical origin of the outer-ring emission, which we addressed in **II-D**.

The ring type of emission emerges around the central main field with increasing pump power due to the spatial hole burning and associated population lens effect discussed in **II-D**. As the pump power increases further, however, the outer-ring emission merges into the central $TEM_{00}$ mode with the increased mode volume as a result of the increased thermal lens effect. A new outer-ring emission then emerges around the merged central mode as a result of the spatial hole-burning and associated population lens effect. This "emerge-merge" process repeats successively up to the maximum absorbed pump power of 300 mW with 9% slope efficiency.

The observed peculiar lasing pattern strongly suggests that a partially modified $TEM_{00}$ mode with structurally unstable outer-ring emissions can maintain itself presumably through nonlinear coherent lateral energy flow from the outer-ring to the single $TEM_{00}$ cavity eigenmode, where the increased thermal lensing dominates the anti-guidance effect and merges the outer-ring emission into the central mode. Note that the video clips clearly show the structural instability featuring the merge-emerge process of the outer-ring emissions with increasing pump power.

An interesting future task might be a quantitative theoretical analysis including thermal and population lens effects responsible for the pump-dependent peculiar lasing pattern formation together with the successive emerge-merge process of outer-ring emissions. In the meantime, the intensity distributions of the near-field patterns shown in Fig. 6 were examined by the PbS infrared viewer followed by the beam analyzer. The results are shown in Fig. 7, which depicts pump-dependent intensity profiles across the vertical axis. It is interesting that the merged intensity distributions approach single-frequency super-Gaussian profiles, $I(r) = I_o \exp\left[-2\left(\frac{r}{w_o}\right)^n\right]$, with steep edges presumably due the coherent superposition of the outer-ring emissions with the central Gaussian profiles.

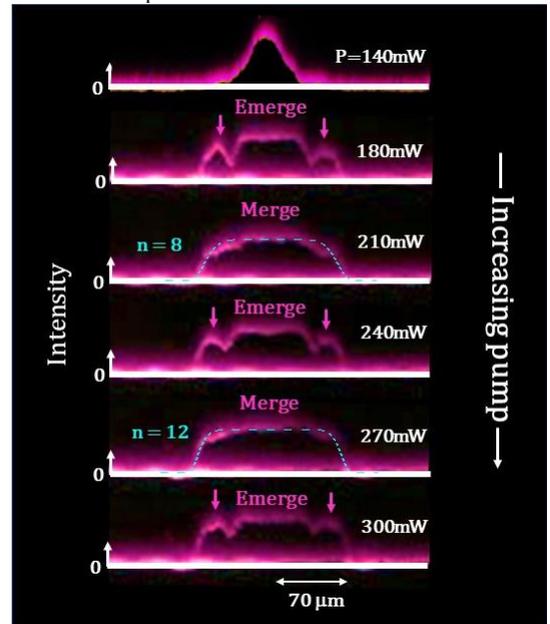

FIG. 7. Pump-dependent intensity distributions of the near-field patterns. Fitting curves are indicated by dashed lines for P = 210 mW and 270 mW.

The dashed lines for P = 210 mW and 270 mW are fitting curves for a lasing spot size of $w_o = 70$ μm and order $n = 8$ and 12. The mode volume, $V_o = \iiint I(x,y,z)dv$, of the super-Gaussian mode with $w_o = w_p = 70$ μm, which is proportional to the output power, is calculated to be twice that of the central Gaussian mode with $w_o \cong 50$ μm in Fig. 5(a). This implies that the modal gain for the modified $TEM_{00}$ is enhanced with the most of the population inversion being utilized for lasing, as discussed in **II-D**. Furthermore, the input-output intensity characteristics shown in Fig. 3(c), which were obtained assuming $w_o = 50$ μm, are preserved at $S_e \cong 21.8\%$ in the super-Gaussian mode with $w_o = 70$ μm because the ratio of the beam cross sections is 1.96.

It is well known that a super-Gaussian beam is not a free-space mode in contrast to a Gaussian beam and the shape of the intensity profile will change during propagation [28]. The steeper the edges of the intensity profile are, the more rapidly will such changes occur. Such a propagation non-invariance of super-Gaussian mode is considered to result in the pronounced shrinkage of central bright spots with respect to the outer-ring diameter in the far-field patterns, in comparison with the near-field patterns, as shown in Fig. 6.

## IV. MODIFIED TEM$_{00}$ MODE OSCILLATIONS TOWARD SELF-MIXING TS$^3$L METROLOGY WITH EXTREME OPTICAL SENSITIVITY

### A. Optical sensitivity versus quantum noise effect toward versatile self-mixing laser metrology

There is growing interest in using TS$^3$Ls for highly sensitive self-mixing metrology owing to their large fluorescence-to-photon lifetime ratio, K, as mentioned in the Introduction. In particular, modified TEM$_{00}$ mode operation with orders of magnitude larger K than that of the normal TEM$_{00}$ mode operation ($1.93 \times 10^7$ versus $10^5$-$10^6$) [8] is expected to provide new insights into ultra-high-sensitivity self-mixing TS$^3$L metrology. In this section, we show experimental results on self-mixing LDV utilizing the modified TEM$_{00}$ mode laser and numerical simulations of semiclassical equations for self-mixing LDV including the effect of spontaneous emission.

The effect of spontaneous emission is crucial for investigating the SNR of self-mixing LDVs in addition to the fluorescence-to-photon lifetime ratio, K. In 1946, E. M. Purcell described that, for a system coupled to an electromagnetic resonator, the spontaneous emission probability is higher than its bulk value and the recombination time is reduced by a factor $F_P = 3\lambda_c^3 Q_c / 2\pi^2 V_c$, which is now called the Purcell factor [16, 29]. Here, $V_c$ is the volume of the resonant mode, $Q_c$ its quality factor, and $\lambda_c = \lambda_o/n_0$ the wavelength in the material.

The spontaneous emission coupling factor, $\beta$, is defined as the fraction of spontaneously emitted photons that are coupled to the cavity mode [29]:

$$\beta = \frac{F_P R_{sp,cav}}{R_{bulk}} \quad (10)$$

where $R_{sp,cav}$ is the rate of spontaneous emissions into the cavity mode per unit time and $R_{bulk}$ is the total spontaneous emission rate per unit time in the bulk medium in the absence of a cavity. $\beta$ becomes equal to $F_P$ in the case of an ideally matched emitter whose volume coincides with the pumped bulk volume, i.e., $R_{sp,cav} = R_{bulk}$. Moreover, it can be substantially increased (decreased) depending on the mode matching between the pump and lasing beam profiles, i.e., $w_p < w_o$ ($w_p > w_o$), as will be discussed later on.

To begin with, let us show the following conventional laser rate equations for population and photon densities without phase equations to better understand the relation between two factors, K and β, in laser dynamics. The self-mixing modulation due to the interference between the lasing field and the coherent component of a feedback field from a target is described by the following rate equations:

$$\frac{dn}{dt} = w - n - ns \quad (11)$$

$$\frac{ds}{dt} = K[\{n - (1 + m\cos(2\pi f_D t))\}s + \beta n]. \quad (12)$$

Here, $w = P/P_{th}$ is the normalized pump power, $n = N/N_{th}$ is the normalized population inversion density, $s = S/\bar{S}(w = 2)$ is the photon density normalized by the steady-state value at $w = 2$, and time is normalized by the fluorescence lifetime, τ. $K = \tau/\tau_p$ is the fluorescence-to-photon lifetime ratio. $m = 2R_f$, $R_f \equiv |E_b/E_f|$ is the amplitude feedback ratio ($E_f$ is the lasing output field amplitude, and $E_b$ is the feedback field amplitude from the moving target), $f_D$ is normalized Doppler-shift frequency and $\beta$ is the spontaneous emission coupling factor described above. It should be noted that large $K$ values in TS$^3$Ls enhance the effective modulation index. Moreover, the effect of quantum (spontaneous emission) noise expressed by the last term of Eq. (12) seems to be enhanced by $\beta K$ in an actual metrology system, as described in the Introduction.

### B. Generalized semiclassical model of self-mixing thin-slice solid-state laser Doppler velocimetry

The rate equations (10) and (11) cannot treat the dynamic effect of spontaneous emission exhibiting Gaussian white noise with zero mean and the value $<\xi(t) \xi(t')> = \delta(t - t')$ is δ-correlated in time.

Thus, the following generalized dynamical equations for the lasing electric field and population inversion, which can be obtained by extending the Lang-Kobayashi equations [30, 31] to include the effect of quantum (spontaneous emission) noise as well as Doppler-shifted delayed feedback of light scattered from a moving target, are employed in the numerical simulations:

$$\frac{dN(t)}{dt} = \frac{2\{w - 1 - N(t) - [1 + 2N(t)]E(t)^2\}}{K} \quad (13)$$

$$\frac{dE(t)}{dt} = N(t)E(t) + R_f E(t - t_D)\cos\Phi(t) + \sqrt{2\beta[N(t) + 1]}\xi(t) \quad (14)$$

$$\frac{d\phi(t)}{dt} = R_f \frac{E(t - t_D)}{E(t)} \sin \Phi(t) \quad (15)$$

$$\Phi(t) = \Omega_D t - \phi(t) + \phi(t - t_D) \quad (16)$$

Here, $E(t) = (g\tau)^{1/2}\mathbf{E}(t)$ is the normalized field amplitude, and $N(t) = g\mathbf{N}_{th}\tau_p(\mathbf{N}(t)/\mathbf{N}_{th} - 1)$ is the normalized excess population inversion, where $\mathbf{N}_{th}$ is the threshold population inversion. $g$ is the differential gain coefficient, where gain is defined as $\mathbf{G} = \mathbf{G}_{th} + g(\mathbf{N}(t) - \mathbf{N}_{th})$. $w = P/P_{th}$ is the relative pump rate normalized by the threshold, $\phi(t)$ is the phase of the lasing field, $\Phi(t)$ is the phase difference between the lasing and the feedback field, and $R_f$ is the amplitude feedback ratio. $\Omega_D = \omega_D/\kappa$ is the normalized instantaneous frequency shift of the feedback light from the lasing frequency. $t$ and $t_D$ are the time and delay time normalized by the damping rate of the optical cavity $\kappa = 1/(2\tau_p)$. The last term of Eq. (14) includes multiplicative quantum (spontaneous emission) noise, where $\beta$ is the spontaneous emission coupling factor.

### C. Laser Doppler velocimetry experiments

The $TEM_{00}$ mode oscillations featuring outer-ring emissions described in **II** possess an extremely short photon lifetime and the resultant fluorescence-to-photon lifetime ratio is very large: $K = 1.93 \times 10^7$. The wide-aperture pumping scheme, $w_p > w_o$, is thought to reduce the spontaneous emission coupling factor, $\beta$. Consequently, highly sensitive laser Doppler velocimetry with a high signal-to-noise ratio (SNR) can be expected.

To verify these speculations, we performed self-mixing modulation experiments. The experiments used Doppler-shifted light scattered from a rotating cylinder toward the TS³L cavity, as depicted in Fig. 1(b), where 95% of the output beam was focused on the cylinder 60 cm away from the laser by a lens with a 15-cm focal length placed 30 cm away from the laser. Here, the LNP laser was modulated at $f_D = 2v/\lambda_o$ because of the self-mixing modulation at the beating frequency between the laser and Doppler-shifted light scattered toward the TS³L cavity ($v$: moving speed along the lasing axis).

First, let us examine the effect of the spontaneous emission coupling factor on LDV signals on the basis of numerical simulations of Eqs. (13)-(16). The results shown in Figs. 8(a) and 8(b) assume parameters relevant to the pure and modified $TEM_{00}$ mode oscillations, respectively. Here, the power spectra were obtained by taking a fast Fourier transformation (FFT) of the calculated time series of 160 μs. As $\beta$ increases, the noise level increases and the SNR of the LDV signal deteriorates accordingly.

Figure 9 shows a typical example of self-mixing laser Doppler velocimetry signals observed for modified $TEM_{00}$ mode operation with $\tau_p = 6.74$ ps corresponding to Fig. 2(c) and 6, together with that for pure $TEM_{00}$ mode operation with $\tau_p = 70.2$ ps corresponding to Fig. 2(a). Here, each power spectrum was obtained by averaging 100 power spectra measured at intervals of the update, 160 μs. The signal broadening at $f_D$ results from both the Gaussian photon statistics of the light fields scattered toward the laser and the fluctuation in the rotation speed of the cylinder. As for the modified $TEM_{00}$ operation with $\tau_p = 6.74$ ps, a variable optical attenuator was inserted before the rotating cylinder. The round-trip attenuation was $T_A = 0$ dB and –15 dB for Fig. 9(a) and –6 dB and –28 dB for Fig. 9(b), respectively. Here, the overall intensity feedback ratio is given by $\eta = 20\log R_f - T_A$ [dB]. It is obvious that the optical sensitivity of the modified $TEM_{00}$ oscillation was 20 dB higher than that of the pure $TEM_{00}$ oscillation, reflecting its extremely large $K$.

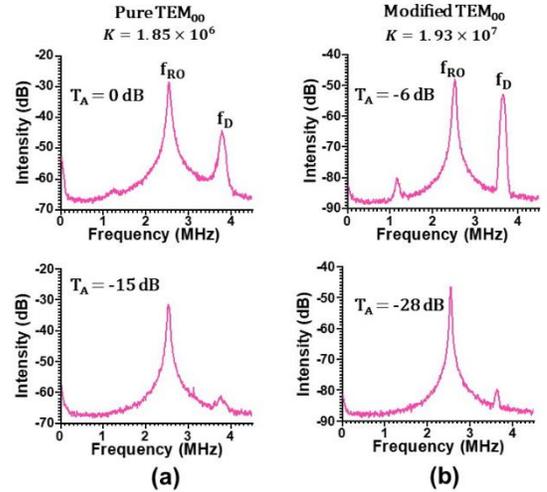

FIG. 9. Observed LDV power spectra for different attenuations, $T_A$, for (a) pure and (b) modified $TEM_{00}$ mode oscillations. P = 65 mW ($w$ = 3.28), (b) 146 mW ($w$ = 1.22). Here, the two peaks at $f_{RO}$ and $f_D$ respectively indicate the relaxation oscillation and Doppler-shift frequencies.

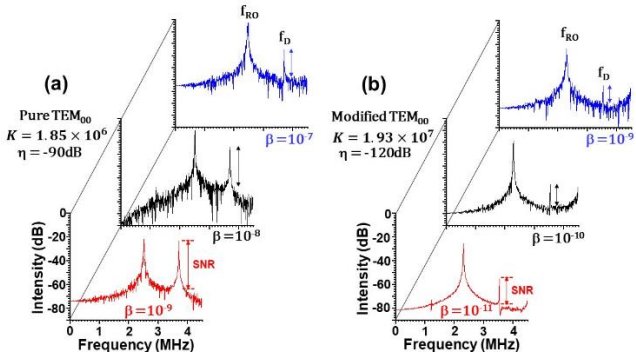

FIG. 8. Dependence of LDV power spectra on spontaneous emission coupling factor, $\beta$. (a) $w$ = 3.28, (b) $w$ = 1.22.

Figures 10 (a) and 10 (b) show example power spectra, indicated by different colors, for different Doppler shift frequencies, $f_D$, where the attenuation was $T_A = 0$ dB and –10 dB for pure and modified $TEM_{00}$ mode operations corresponding to Fig. 2(a) and 2(c), respectively. An FFT was carried out on the temporal evolution of the output waveform during the observation period of 160 μs. The SNR of the LDV signals is plotted as a function of $f_D$ in Fig. 10(c). The measured SNR for the modified $TEM_{00}$ mode oscillation with $T_A = 10$ dB is 10 dB larger than that for the pure $TEM_{00}$ mode oscillations without attenuation, i.e., $T_A = 0$ dB. This implies that the SNR can be improved by 20 dB at the same intensity feedback ratio from the target for the modified $TEM_{00}$ mode oscillations.

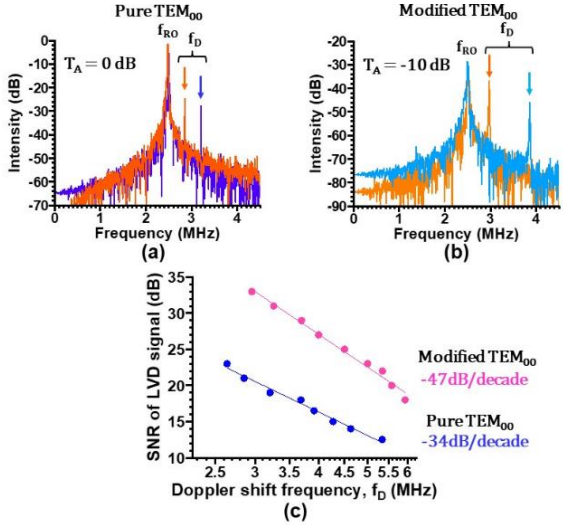

FIG. 10. (a), (b) LDV signals for different $f_D$ values. (c) Signal-to-noise ratio versus Doppler-shift frequency, $f_D$, measured for pure and modified $TEM_{00}$ mode operation.

Numerical simulations were performed to estimate the actual feedback ratio of the present LDV scheme. The power spectrum in the absence of the optical attenuator, $T_A = 0$, shown in Fig. 9(a) was well reproduced by assuming an intensity feedback ratio of $\eta = -94$ dB. Here, the power spectra were obtained by performing an FFT on the calculated time series of 160 μs for the $TEM_{00}$ mode oscillations with $\tau_p = 70.2$ ps, assuming $\beta = 4.84 \times 10^{-8}$. In short, the effective intensity feedback ratio solely from the rotating cylinder was very small, $-94$ dB.

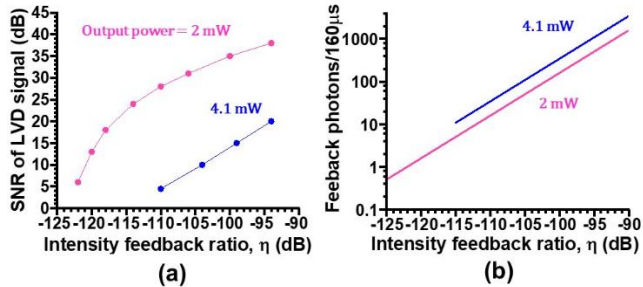

FIG. 11. (a) SNR of LDV signals measured as a function of the intensity feedback ratio, η, which was calibrated on the basis of a careful numerical fitting. (b) Average number of feedback photons per observation time.

The experimental SNRs of the LDV signals for the pure and modified $TEM_{00}$ mode oscillations are plotted in Figs. 11 (a) as a function of the overall intensity feedback ratio, η, by changing the $T_A$ value of the attenuator and assuming that the intensity feedback ratio is solely from the rotating cylinder without the attenuator, $\eta = -94$ dB.

The numerically calculated LDV power spectra for pure $TEM_{00}$ and modified $TEM_{00}$ modes are shown in Figs. 12 (a) for $\beta = 4.84 \times 10^{-8}$ and (b) for $\beta = 1 \times 10^{-9}$. The numerical results reproduce experimental SNR values shown in Fig. 11 (a) quite well. In the range $\eta < -110$ dB, the measured SNR for the modified $TEM_{00}$ mode operation at an output power of 2 mW tends to deviate from the linear relation expected for stationary LDV. This point will be discussed in the section **D**. The dynamic range of the LDV measurement is 40 dB for modified and 20 dB for pure $TEM_{00}$ operation.

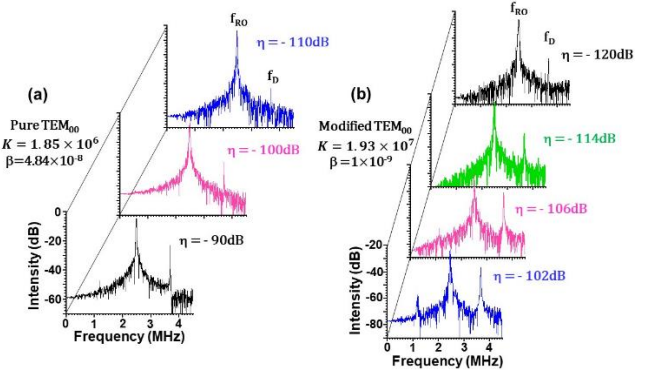

FIG. 12. Numerical LDV power spectra for different intensity feedback ratios, η. (a) $w = 3.28$, (b) $w = 1.22$.

The modified $TEM_{00}$ oscillations became chaotic as the optical attenuation was decreased, although the light scattering coefficient of the rotating cylinder toward the laser was quite small. Figure 13(a) shows a typical example of chaotic spiking waveforms indicating a "subharmonic resonance" at $f_{SP} = f_D/2$ [31], together with the corresponding power spectrum, and Fig. 13(b) shows numerical results exhibiting a subharmonic resonance nature for $\eta = -94$ dB.

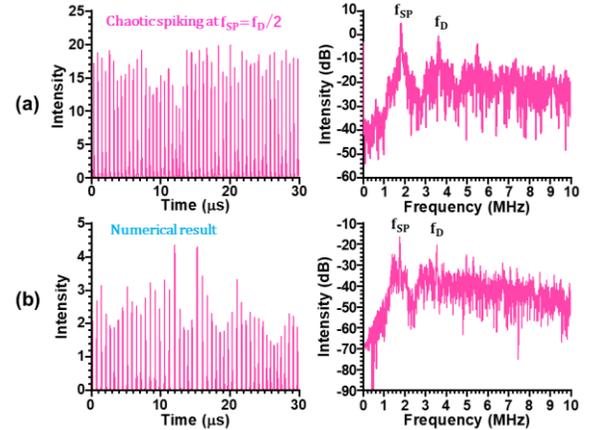

FIG. 13. Spiking chaos featuring subharmonic resonance.

### D. Effect of number of self-mixing photons

The optical sensitivity of self-mixing TS$^3$L metrology depends on the fluorescence-to-photon lifetime ratio as well as the spontaneous emission coupling factor. In addition, the sensitivity is considered to be limited by the number of self-mixing photons fed back to the TS$^3$L in the extremely small η regime.

Figure 11(b) plots the number of photons fed back during the observation period, $N_p$, that induce self-mixing laser modulation through interference with the lasing field in the cavity, for the pure and modified TEM$_{00}$ mode operations in Fig. 11(a). Here, the observation period was 160 μs and output laser power was 4.1 mW or 2 mW, like the experimental conditions. In the range η < -110 dB, the average number of self-mixing photons during the observation period fell below $N_p$ < 10. In this situation, "intermittent" LDV modulation might take place for each power spectrum measurement over the minimum update interval of 160 μs because only a few Doppler-shifted photons are fed back to the TS$^3$L, for which the arrival times might obey Poisson statistics [32]. The excess reduction in SNR of LDV signals, i.e., the deviation from the linear SNR versus η relation of the stationary LDV modulation, shown in Fig. 11(a) is thought to be due to the intermittent modulation. Previously, we demonstrated similar self-mixing LDV measurements with few feedback photons per observation period in a dual-polarization Nd:GdVO$_4$ TS$^3$L, operating in annular near-field and super-Gaussian far-field patterns, with a 1-mm-thick platelet cavity tilted slightly from the LD pump beam direction [33].

Although a full quantum mechanical treatment of single photon self-mixing modulation remains to be done, one can imagine the following thought experiment: *What does the model composed of the laser field equations (13)-(16) look like when multiplicative quantum (spontaneous emission) noise is included?*

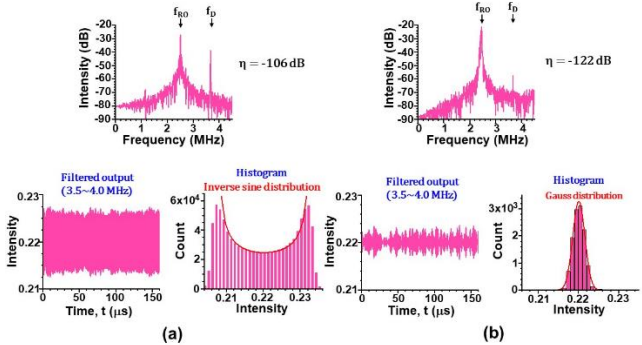

FIG. 14. Numerical results for the modified TEM$_{00}$ mode oscillation at different intensity feedback ratios. $w = 1.22$.

Figures 14(a) and 14(b) show numerical results for the modified TEM$_{00}$ operation, including the power spectra of the calculated intensity time series with a 160 μs period, filtered intensity waveforms through 3.5 – 4 MHz and their intensity probability distributions for different intensity feedback ratios. Stationary LDV signals at $f_D$ appear in the large η regime, and the resultant intensity histogram nearly follows an inverse sine distribution. As η decreases into the regime of single or a few feedback photons, non-stationary LDV signals appear that are strongly perturbed by quantum (spontaneous emission) noise; in particular, there are irregular "breathing" oscillations where the upper and lower envelopes of beat oscillating at $f_D$ exhibit a symmetric variation with respect to the average value. The resultant intensity histogram has a Gaussian distribution. Such a breathing phenomenon would decrease the SNR of the LDV signals even in the case of model laser field equations (13)-(16).

To provide further insights into the irregular "breathing" beat-wave oscillations, statistical analyses were performed on the plots in Fig. 14. Temporal evolutions of beat signal amplitude are shown in Figs. 15, together with their envelope variations. Here, the amplitudes of the beat wave oscillations, which is defined as $\triangle E(t) \equiv E_u(t) - E_l(t)$, are plotted. For a large intensity feedback ratio, the beat-wave amplitude probability distribution follows a Gaussian function (Fig. 15(a)). Here, the beat-wave amplitude scales as $\triangle E(t) = w_B \times 10^3$, where the bin width for the histogram analysis is $w_B = 1 \times 10^{-3}$.

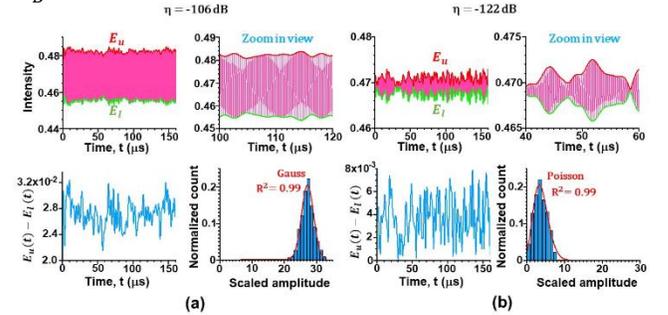

FIG. 15. Numerical analysis of breathing beat-wave oscillations of the modified TEM$_{00}$ mode, indicating (a) Gauss statistics changing to (b) Poisson statistics with decreasing intensity feedback ratio.

It is surprising that the histogram in Fig. 15(b) appears to obey a Poisson function with a high coefficient of determination, $R^2 = 0.99$, when the factorial term, n!, in the Poisson function is replaced with a gamma function, $\Gamma(n+1)$, so that all positive real numbers in the numerical time series can be fitted. The physical meaning of the scaling has yet to be understood, but it is conjectured that there is a correspondence between the beat-wave amplitude and the number of photons involved in the self-mixing modulations.

Figure 15 indicates that the probability distributions of the amplitude of beat-wave oscillations, which were found in this study, are Poisson when the intensity of the feedback light decreases. This presents an intriguing analogy to the photon statistics in quantum optics wherein the Gaussian distribution approaches a Poisson distribution as the number of photons decreases, although the dual wave-particle nature of the self-mixing photons is not incorporated in the present numerical experiment.

Finally, let us examine some typical experimental self-mixing LDV results for the modified TEM$_{00}$ mode operation in the regime of a few self-mixing photons per observation period. As shown in Fig. 16, the experimental results are similar to the Gaussian histogram for the "breathing" beat-wave oscillations (Fig. 14(b)). Moreover, the histogram for the scaled beat-wave amplitude apparently obeys a Poisson distribution with $R^2 = 0.96$, like the numerical result in Fig. 15(b).

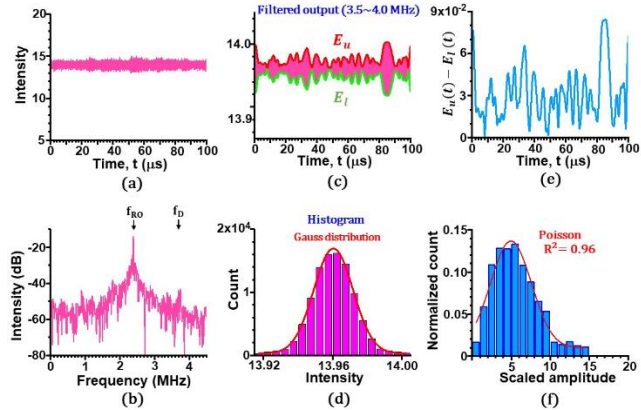

FIG. 16. Experimental results showing self-mixing LDV in the regime of few feedback photons per 100-μs observation period. P = 140 mW, η = −120 dB. (a), (b): LDV intensity waveform and its power spectrum. (c), (d): Filtered waveform and its intensity probability distribution. (e), (f): Beat-wave amplitude fluctuation and corresponding histogram. The histogram assumes $\triangle E(t) = w_B \times 250$.

In the self-mixing LDV experiment with the rotating cylinder, the stability of the LDV signals seems to be restricted by the unevenness of the rough-metal cylinder. More light could be shed on the situation by conducting systematic experimental studies in a wide parameter region utilizing highly stable frequency-shifted optical feedback with an acousto-optic modulator [34] instead of a rotating cylinder, in conjunction with accurate counting of the feedback photons by using a single-photon detector.

We believe that such a quantum-classical correspondence could provide new physical insights into self-mixing TS$^3$L metrology systems operating with much less than one feedback photon per modulation cycle [33, 35].

## V. CONCLUSIONS

A peculiar transverse mode oscillation, i.e., a TEM$_{00}$ mode accompanied by an outer-ring emission, possessing an extremely short photon lifetime was observed in a 300-μm-thick platelet LiNdP$_4$O$_{12}$ (LNP) laser with wide-aperture laser-diode end pumping.

Precise analyses revealed that the modified TEM$_{00}$ mode manifests itself in the transverse lasing pattern in which most of the population inversions take part in lasing through the combined effect of transverse spatial hole-burning and thermal lensing and they compensate for the large cavity loss resulting from the anti-guidance due to the population lens effect. The structural instability, featuring the merge-emerge process of outer-ring emissions with increasing pump power, can be interpreted in terms of competing thermal and population lens effects. In addition, the modified TEM$_{00}$ laser was shown to possess an extremely short photon lifetime of $\tau_p = 6.74$ ps.

We demonstrated self-mixing laser Doppler velocimetry with an LNP laser operating in modified TEM$_{00}$ mode. A 20-dB enhancement in SNR was achieved in comparison with the pure TEM$_{00}$ mode operation, reflecting the decreased spontaneous emission coupling factor of $\beta = 10^{-9}$ associated with the wide-aperture pumping, together with the increased K value. The results of the self-mixing LDV experiments were numerically verified using a model of a single-mode laser subjected to external frequency-shifted feedback and multiplicative quantum (spontaneous emission) noise.

This enabled us to achieve a high signal-to-noise ratio with the LDV signals in the regime of few feedback photons per observation period. Moreover, the following intriguing analogy to the photon statistics in quantum optics was revealed by the numerical simulations and experiments: the amplitude of the breathing beat wave at the Doppler-shifted frequency, which was found in this study, eventually obeys Poisson statistics as the number of self-mixing photons is decreased, whereas it approaches Gaussian statistics as the number of self-mixing photons increases.